\documentstyle[12pt]{article}
\newcommand{\f}{\begin{equation}}
\newcommand{\e}{\end{equation}}
\newcommand{\n}{\label}
\begin{document}
\title{ TWISTORS AND NEARLY AUTOPARALLEL MAPS}
\author{{\it Sergiu I. Vacaru}\thanks{E-mail: lises@cc.acad.md } \\
 Institute of Applied Physics,\\  Academy of Sciences of Moldova, \\
 5 Academy str., Chi\c sin\v au, 2028, Republic of Moldova \\
and \\
 {\it Sergiu V. Ostaf} \\
 Department of Mathematics and Physics, \\ Tiraspol State University
(evacuated in Chi\c sin\v au), \\
 5 Iablochkin str., Chi\c sin\v au, 2062, Republic of Moldova}
\maketitle
\begin{abstract}
The purpose of the present paper is to investigate the problem
of definition of twistors on generic curved spaces. Firstly, we consider
nearly geodesic ( autoparallel ) maps of (pseudo)--Riemannian spaces .
Secondly, we shall define nearly autoparallel twistor equations which are
compatible
on nearly conformally flat spaces. Finally, we shall study nearly autoparallel
twistor structures generating curved spaces and vacuum  Einstein
spaces.
\end{abstract}

\section{Introduction: Spinors and maps of curved spaces with deformation
of connections}

Our geometrical constructions will be realized on pairs of 4--dimensional
(pseudo)--Riemannian spaces $(V, {\underline V})$~ with signature $(- + + +)$~
and 1--1 local maps of spaces $ f : V {\to} {\underline V}$, given by functions
$ f^{\underline {\mu}}(x)$~ of smoothly class ${\cal C}^{r}(U), (r>2, r={\omega}$~
for analytic functions ) and theirs inverse functions $f^{{\mu}}({\underline x})$~
with corresponding non--zero  Jacobians in every  point
$x=\{x^ {\mu}\} {\subset} U {\subset} V$
and
${\underline x}=\{ {\underline x}^{\mu}\} {\subset}
{\underline U} {\subset} {\underline V} \quad  (U$~ and ${\underline U}$~
are open
regions ). We shall attribute  regions $U $~ and ${\underline U}$~
to a common for a given f--map coordinate system when every point $q{\in}U$, ~
with coordinates $x^{\mu}(q)$, is mapped into a point
${\underline q} {\subset} U$~
with the same  coordinates $x^{\mu}={\underline x}^{\mu}({\underline q})=
x^{{\underline {\mu}}}({\underline q}),$ i.e.
$ f : x^{\mu}(q) \to x^{\mu}({\underline q}).$
We note that all calculations in this work will be local in nature
and taken to refer to common coordinates, for given f--maps, on open regions
of spaces into considerations.

The metric tensor, con\-nec\-tion, dif\-fer\-en\-tial
 op\-er\-a\-tor and tetrads (frames) on $U$~ and
 ${\underline U}$ are correspondingly denoted as
$ g_{{\mu} {\nu}}(x),
 {\Gamma}^{{..}{\alpha}}_{{\mu}{\nu}}(x)$, $D_{\mu} $ and
$h^{a}_{{\mu}}(x),$ where
\f
 g_{{\mu}{\nu}}(x)=h^{a}_{\mu}(x) h^{b}_{\nu}(x){\eta}_{ab},
\quad {\eta}_{ab}=const,
 \n{f1}
\e
and, respectively, $ {\underline g}_{{\mu}{\nu}}(x){\equiv}
g_{{\underline {\mu}}{\underline {\nu}}}(x),\
{\underline {\Gamma}}^{{..}{\alpha}}_{{\mu}{\nu}}(x){\equiv}
{\Gamma}^{{..}{{\underline \alpha}}}_{{\underline {\mu}}{\underline {\nu}}}(x),
{\underline D}_{\mu}=D_{{\underline \mu}}$
and ${\underline h}^{a}_{\mu}(x)=h^{a}_{\underline {\mu}}(x),$  where
 \f {\underline g}_{{\mu}{\nu}}(x)={\underline h}^{a}
_{\mu}(x){\underline h}^{b}_{\nu}(x){\eta}_{ab}, \quad
  {\eta}_{ab}=const  \n{f2} \e
(Greek indices ${\mu}, {\nu}, {\underline {\mu}}, {\underline {\nu}},{\ldots}$~
run from $0$~ to $3$~).
Throughout the present paper we shall use the
terminology and definitions of abstract
and coordinate tensor and spinor index formalisms described in monographs
 [1, 2].
For simplicity,  we shall write  Latin symbols
$ a, b, {\underline a}, {\underline b}, {\ldots}$  for both abstract and  tetrad
indices implying that in the first case Latin indices are abstract labels
and in the second case (for decompositions on tetrads )
we shall specify their explicit values $(a, b, {\ldots}=0, 1, 2, 3).$
We consider spinor decompositions of metrics (1) and (2) :
\f
g_{{\mu}{\nu}}={\sigma}_{\mu}^{AA^{\prime}}(x) {\sigma}_{\nu}^{BB^{\prime}}(x)
{\epsilon}_{AB}{\epsilon}_{ A^{\prime} B^{\prime}},  \n{f3}
\e
where ${\sigma}_{\mu}^{AA^{\prime}}(x)=h^{a}_{\mu}(x){\sigma}_{a}^{AA^{\prime}},
{\sigma}_{a}^{AA^{\prime}}=const, $ are Infeld--van der Waerden coefficients
and ${\epsilon}_{AB}=-{\epsilon}_{BA}, {\epsilon}_{A^{\prime}B^{\prime}}=
-{\epsilon}_{B^{\prime}A^{\prime}}  (A, B, A^{\prime}, B^{\prime}=0, 1)$~
are spinor metrics
\f
 ( {\underline g}_{{\mu}{\nu}}={\underline {\sigma}}_{{\mu}}^
{AA^{\prime}}(x){\underline {\sigma}}_{\nu}^{BB^{\prime}}(x)
{\epsilon}_{AB}{\epsilon}_{A^{\prime}B^{\prime}}, \n{f4}  \e
where ${\underline {\sigma}}_{\mu}^{AA^{\prime}}(x)=
{\sigma}_{\underline {\mu}}^{AA^{\prime}}(x)=
{\underline h}^{a}_{\mu}(x){\sigma}_{a}^{AA^{\prime}};$
if necessary we shall write, for example, ${\underline {\epsilon}}_{AB},
{\underline {\epsilon}}_{A^{\prime}B^{\prime}}, {\underline {\sigma}}_{a}^
{A,A^{\prime}}, {\underline {\omega}}^{A}$~ or ${{\underline {\xi}}}^{AB}_{A^{\prime}}$
in order to point that these spin--tensor values are associated to the spinor
decomposition (4) on space ${\underline V}$).

For mutual transformations of tensor and spinor indices one introduces
inverse Infeld--van der Waerden coefficients  ${\sigma}^{\mu}_{AA^{\prime}}(x)$~
and ${\underline {\sigma}}^{\mu}_{AA^{\prime}}(x)$, for
example, $B^{\alpha}={\sigma}^{\alpha}_{AA^{\prime}}B^{AA^{\prime}}$~ and
${\underline A}_{BB^{\prime}}=A_{\underline {\mu}}{\sigma}^{\underline {\mu}}_
{BB^{\prime}}.$

Covariant derivation of spinors on $V$~ is defined by using spin coefficients
${\gamma}_{AA^{\prime}B}^{{...}C}$ and
${\gamma}_{AA^{\prime}B^{\prime}}^{{...}C^{\prime}}:$
$$D_{AA^{\prime}}{\xi}^{B}=
{\sigma}^{\mu}_{AA^{\prime}}(x)D_{\mu}{\xi}^{B}=
{\partial}_{AA^{\prime}}{\xi}^{B}+
{\gamma}_{AA^{\prime}{C}}^{{...}{B}}{\xi}^{B}$$
and $ D_{AA^{\prime}}{\xi}_{B^{\prime}}=
{\sigma}^{\mu}_{AA^{\prime}}(x)D_{\mu}{\xi}_{B^{\prime}}=
{\partial}_{AA^{\prime}}{\xi}_{B^{\prime}}-
{\gamma}_{AA^{\prime}B^{\prime}}^{{...}C^{\prime}}{\xi}_{C^{\prime}}.$

In a similar manner spin coefficients ${\underline {\gamma}}_
{AA^{\prime}{B}}^{{...}{C}}$
and ${\underline {\gamma}}_{AA^{\prime}B^{\prime}}^{{...}C^{\prime}}$
are considered in order to define covariant derivation of spinors
 on ${\underline V}$.

Here we present formulas interrelating spin coefficients ${\gamma}_{{\mu}B}^
{{..}{C}}$ and ${\gamma}_{{\mu}B^{\prime}}^{{..}C^{\prime}}$
with Christoffel symbols in the case of orthonormalized bases (when
$ {\epsilon}_{A^{\prime}B^{\prime}}= {\epsilon}_{AB}
=\left(\matrix{0&1\cr
        -1&0\cr}\right)):$
\f {\gamma}_{{\mu}{B}}^{{..}{C}}
={\Gamma}_{{\mu}{B}C^{\prime}}^{{...}{C}C^{\prime}}-
{1 \over 2}{\sigma}^{\nu}_{{B}B^{\prime}}{\partial}_{\mu}
{\sigma}_{\nu}^{{C}B^{\prime}}, \quad
 {\gamma}_{{\mu}B^{\prime }}^{{..}C^{\prime}}=
{\Gamma}_{{\mu}{B}B^{\prime}}^{{...}{B}C^{\prime}}-
{1\over 2}{\sigma}^{\nu}_{{B}B^{\prime}}{\partial}_{\mu}{\sigma}
_{\nu}^{BC^{\prime}}  \n{f5} \e
$$ ({\gamma}_{\underline {\mu}B}^{{..}{C}}=
{\Gamma}_{{\underline {\mu}}BC^{\prime}}^{{...}{C}C^{\prime}}-
{1 \over2}{\sigma}^{\underline {\nu}}_{BB^{\prime}}
{\partial}_{\underline {\mu}}{\sigma}_
{\underline {\nu}}^{{.}{C}B^{\prime}}, \quad
 {\gamma}_{{\underline {\mu}}B^{\prime}}^{{..}{\underline C}}=
{\Gamma}_{{\underline {\mu}}BB^{\prime}}^{{...}{B}C^{\prime}}-
{1 \over 2}{\sigma}^{\underline {\nu}}_{BB^{\prime}}{\partial}_
{\underline {\mu}}{\sigma}_{\underline {\nu}}^{{.}{B}C^{\prime}}).$$
and, inversely, for ${\sigma}_{a}^{AA^{\prime}}=const,$
$$ {\Gamma}_{{B}B^{\prime}{C}C^{\prime}}^{{..}{A}A^{\prime}}=
{\gamma}_{BB^{\prime}C}^{{...}{A}}{\epsilon}_{C^{\prime}}^{{.}A^{\prime}}+
{\overline {\gamma}}_{BB^{\prime}C^{\prime}}^{{...}A^{\prime}}{\epsilon}^
{{.}{A}}_{C} \quad
({\underline {\Gamma}}_{BB^{\prime}CC^{\prime}}^{{....}{A}A^{\prime}}=
{\underline {\gamma}}_{BB^{\prime}C}^{{...}{A}}{\underline {\epsilon}}
_{C^{\prime}}^
{A^{\prime}}+{\overline {\gamma}}_{BB^{\prime}{C^{\prime} }}^{{...}A^{\prime}}
{\epsilon}^{{.}{A}}_{C})$$
where $ {\epsilon}^{{.}{B}}_{A}={\delta}_{A}^{B}, {\epsilon}_{C^{\prime}}^
{B^{\prime}}={\delta}_{A}^{B}$
and ${\overline {\gamma}}_{BB^{\prime}C^{\prime}}^{{...}{A}}$
 denotes complex conjugation of ${\gamma}_{BB^{\prime}C^{\prime}}^{A}).$

Our first objective in this paper is to study deformations of spinor objects
in result of superposition of local 1--1 maps $f: V {\to} {\underline V}$
with deformation of connection
\f
{\Gamma}^{{..}{\alpha}}_{{\beta}{\gamma}}(x)=
{\Gamma}_{{\beta}{\gamma}}^{{..}{\alpha}}(x)+
P^{{..}{\alpha}}_{{\beta}{\gamma}}(x),     \n{f6}
\e
and, in consequence of formulas (5), deformations of spin coefficients, for example,
\f {\underline {\gamma}}_{{\mu}B}^{{..}{C}}(x)=
{\gamma}_{{\mu}{B}}^{{..}{C}}(x)+
{^{\star}{\gamma}}_{{\mu}{B}}^{C}(x)  \n{f7} \e
where $P^{{..}{\alpha}}_{{\beta}{\gamma}}(x)$
and ${^{\star}{\gamma}}^{{..}{C}}_{{\mu}{B}}=
P_{{\mu}BC^{\prime}}^{{...}{C}C^{\prime}}-
{1 \over 2}({\sigma}^{\underline {\nu}}_{BB^{\prime}}{\partial}_{\mu}
{\sigma}^{CB^{\prime}}_{\underline {\nu}}-
{\sigma}_{BB^{\prime}}^{\nu}{\partial}_{\mu}{\sigma}_{\nu}^{CB^{\prime}})$
are called the deformation tensor and, respectively, the deformation spin tensor.
Deformations of the covariant derivation operator,
 cased by splittings of type (6),
 or (7),
will be denoted as
\f
{\underline D}_{\mu}=D_{\mu}+{^{\star}D}_{\mu}, \quad
{\underline D}_{AA^{\prime}} = D_{AA^{\prime}} +
{^{\star}D}_{AA^{\prime}}  \n{f8} \e

In a particular case of conformal maps
$c :U {\to} {\underline U}$,  when
$${\underline g}_{ab}={\Omega}^{2}g_{ab}, \  {\underline {\epsilon}}_{AB}=
{\Omega}{\epsilon}_{AB}, \ {\underline {\epsilon}}_{A^{\prime}B^{\prime}}=
{\Omega}{\epsilon}_{A^{\prime}B^{\prime}},$$
${\Omega}(x)$ is  a nonzero real function on $U$,
\f {^{\star}D}_{\mu}={\Omega}^{-1}D_{\mu}{\Omega}=D_{\mu}{\ln{\Omega}}.
\n{f9}  \e
Conformal transforms are largely used, for example, in  twistor [2] and
conformal field theories.

We note that there are classes of 1--1 local maps with deformation of
connection ((6) or (7)) more general then that for conformal maps
(9) (see, for example, $(n-2)$--projective spaces [3], nearly geodesic maps,
ng--maps [4], nearly autoparallel maps, na--maps, of spaces with torsion
and nonmetricity [5, 6], of fiber bundles [7] and of Finsler and Lagrange
 spaces [8] ).  In works [5, 9, 10, 11 ]
we have proposed to apply ng-- and na--maps for definition of conservation laws
on curved spaces. Na--maps were used for definition of nearly autoparallel
twistors in connection to a possible twistor--gauge interpretation of vacuum
gravitational fields [7, 11--14].

The second objective is the investigation of na--map deformations of twistor
equations    [2] (for our purposes written on space ${\underline V}$)
\f {\underline D}_{A^{\prime}}^{({A}}{\underline {\omega}}^{{B})}=
{1 \over 2}({\underline D}^{A}_{A^{\prime}}{\underline {\omega}}^{B}+
{\underline D}^{B}_{A^{\prime}}{\underline {\omega}}^{A})=0, \n{f11} \e
where $(\quad )$ denotes symmetrization.

Because for uncharged twistors
\f {\underline {D}}^{A^{\prime}(C}{\underline D}^{A}_{A^{\prime}}
{\underline {\omega}}^{B)}=-
{\underline {\Psi}}^{{CA}{.}{B}}_{{..}{D}}{\underline {\omega}}^{D},
 \n{12}  \e
where
\f {\underline C}_{abcd}{\equiv}{\underline {\Psi}}_{ABCD}
{\underline {\epsilon}}_{A^{\prime}B^{\prime}}{\underline {\epsilon}}_
{C^{\prime}{D^{\prime}}}+
{\underline {\overline {\Psi}}}_{A^{\prime}B^{\prime}C^{\prime}D^{\prime}}
{\underline {\epsilon}}_{AB}{\underline {\epsilon}}_{CD}  \n{f12} \e
is the conformal Weyl tensor on space ${\underline V}$, there is a hard
compatibility condition for twistor equations      (10), namely,
${\underline {\Psi}}_{ABCD}{\underline {\omega}}^{D}=0, $
which characterizes, for example, conformally flat spaces.
That is why a rigorous mathematical, generally accepted, definition of twistors
was possible only for conformally flat spaces and this fact is the main
impediment to a twistor interpretation of general gravitational fields
( see details in [2] ).

Our main idea [11, 7, 13, 14] was to define twistors not on generic curved spaces
$V$, where twistor equations are incompatible,  but to remove the problem
on auxiliary conformally flat (or more simply, flat ${\underline M}$)
 background
spaces ${\underline V}$, interrelated with the fundamental space--time
$V$ by means of chains of na--maps (nearly conformal maps, nc--maps).
On space ${\underline M}$ twistor equations (10) become compatible ;
we can define twistors in a standard manner as pairs of spinors,
${\underline Z}^{\alpha}=({\underline {\omega}}^{A}, {\underline {\pi}}_{A}).$
Then, transferring ${\underline Z}^{\alpha}$ on $V$, by using nc--maps,
we can define nearly autoparallel twistors, na--twistors, as satisfying
na--twistor equations, being na--images of equations (10).
For simplicity , in this paper we shall restrict ourselves only with
nearly geodesically         flat, ng--flat, spaces
$V$,     which admit ng--maps to Minkowski space ${\underline M}$.
We shall analyze conditions when na--twistor equations contain information
on vacuum Einstein fields.

\section{ Nearly geodesic maps and spinors}

The aim of this section is to present a brief introduction into the geometry
of ng--flat spaces. We shall specify basic ng--map equations and invariant
conditions [4] to the case of vacuum gravitational fields on $V$.
Proofs are mechanical, but, in most cases, rather tedious calculations,
similar to those presented in  [4, 13]. They are omitted.
\vskip 6pt

{\it 2.1. Definition of ng--maps.} Let parameterize  curves on
$U {\subset} V$ by functions $x^{\mu}=x^{\mu}(\eta),  {\eta}_{1} <{\eta} <
{\eta}_{2}$, with corresponding tangent vector field defined as
$u^{\mu}={{dx^{\mu}} \over {d{\eta}}}$.
\vskip 6pt

{\sl Definition 1.}{\it A curve $l$ is called a geodesic on $V$ if its tangent
vector field satisfies autoparallel, a--parallel, equations :
$$uDu^{\alpha}=u^{\beta}D_{\beta}u^{\alpha}={\rho}({\eta})u^{\alpha},
 \eqno (13)$$
where ${\rho}({\eta}) $~ is a scalar function}.
\vskip 6pt

We note that for (pseudo)--Riemannian spaces the extremal curves,
the geodesics,
coincide with the straightest curves, a--parallels, and that is why we shall use
term geodesics for both classes of curves (for spaces with locally isotropic
or anisotropic torsion and nonmetricity we have started with a--parallel
equations  [5 -- 8] ).

On space ${\underline V}$ we consider a new class of curves : Let curve
${\underline l} {\subset} {\underline V}$ is given parametrically as
$x^{\alpha} =x^{\alpha} ({\eta}), {\eta}_{1} < {\eta} < {\eta}_{2},
u^{\alpha}={{dx^{\alpha}} \over {d{\eta}}} {\neq} 0.$
We say that a 2--dimensional distribution $E_{2}({\underline l})$~ is
coplanar along ${\underline l}$ if in every point $x{\in} {\underline l}$
it is defined a 2--dimensional vector space
$E_{2}(x) {\subset} T_{x}{\underline V} \quad
 (T_{x}{\underline V}$ is the tangent space to
$x {\in} {\underline V}$) and every vector
${\underline p}^{\mu}(x^{\nu}_{(0)}) {\subset} E_{2}({\underline l}),
x^{\mu}_{(0)} {\in}{\underline l}, $ rests contained in the same distribution
after parallel transports along ${\underline l}$, i.e.
$p^{\alpha}(x^{\beta}
({\eta})) {\subset} E_{2} ({\underline l})$.
\vskip 6pt

{\sl Definition 2.} {\it A curve ${\underline l}$ is called a nearly geodesic
on space ${\underline V}$ if a coplanar along ${\underline l}$
distribution   $E_{2}({\underline l})  $
containing the tangent to ${\underline l}$
vector field $u^{\alpha}({\eta})$ is defined.}
\vskip 6pt

Ng--maps are introduced [4] according the
\vskip 6pt

{\sl Definition 3} {\it Nearly geodesic maps, ng--maps, are local 1--1 mappings
of (pseudo)--Riemannian spaces, $ng : V {\to} {\underline V}, $
changing every geodesic $l$ on $V$ into nearly geodesic ${\underline l}$
on ${\underline V}$ .}
\vskip 6pt

Let a geodesic $l {\subset} U$ is given by functions
 $x^{\alpha}=x^{\alpha}({\eta}),
u^{\alpha}={{dx^{\alpha}} \over {d{\eta}}}, {\eta}_{1} < {\eta} <{\eta}_{2}, $
satisfying equations (12). We suppose that to the geodesic
$l$ corresponds a nearly geodesic ${\underline l} {\subset} {\underline U}$
given by the same parametrization in a common for a chosen ng--map local
coordinate system  on $U$ and ${\underline U}$. This requirement
is satisfied if and only if vectors $u^{\alpha},
{\underline u}^{\alpha}_{(1)}=u{\underline D}u^{\alpha}$
and ${\underline u}^{\alpha}_{(2)}=u{\underline D}u^{\alpha}_{(1)}$
are linearly depended in every point $x{\in} {\underline U},$ i.e.
$$ {\underline u}^{\alpha}_{(2)}={\underline a}({\eta})u^{\alpha}+
{\underline b}({\eta}){\underline u}^{\alpha}_{(1)}$$
for some scalar functions ${\underline a}({\eta})$ and ${\underline b}({\eta}).$
Putting splitting (6) into expressions  for ${\underline u}^{\alpha}_{(1)}$
and ${\underline u}^{\alpha}_{(2)}$ on $U$ and
from the just presented linear
 dependence we obtain :
$$ u^{\beta}u^{\gamma}u^{\delta}(D_{\beta}P^{{..}{\alpha}}_{{\gamma}{\delta}}
+P^{{..}{\alpha}}_{{\beta}{\gamma}}P^{{..}{\tau}}_{{\gamma}{\delta}})=
bu^{\gamma}u^{\delta}P^{{..}{\alpha}}_{{\gamma}{\delta}}+au^{\alpha},
 \eqno (14)$$
where $ b({\eta}, u)={\underline b}-3{\rho}$ and
$$a({\eta}, u)={\underline a}+{\underline b}{\rho}-u^{b}{\partial}_{b}{\rho}-
{\rho}_{2} \eqno (15)$$
are called the  deformation parameters of ng--maps.
\vskip 6pt

{\it 2.2. Classification of ng--maps.} Ng--maps were classed [4] by considering
possible polynomial dependencies on $u^{\alpha}$ of deformation parameters
(15). We shall consider maps $ng: V {\to} {\underline V}$ satisfying reciprocity
conditions (when $ {ng}^{-1} : {\underline V} {\to} V$ is also an ng--map) .
This requirement is fulfilled if
$$ P_{{\alpha}({\beta}}^{{..}{\tau}}P_{{\gamma}{\delta})}^{{..}{\alpha}}=
d_{({\beta}}P^{{..}{\tau}}_{{\gamma}{\delta})}+
c_{({\alpha}{\beta}}{\delta}^{\tau}_{{\gamma})}, $$
for a vector, $ d_{\beta},$ and tensor, $c_{{\alpha}{\beta}},$ on $V$.
\vskip 6pt

{\sl Theorem 1} {\it Four classes of ng--maps  are characterized by corresponding
parametrizations of deformation tensors and basic equations :

--for trivial ng--maps, geodesic maps (or ${\pi}_{(0)}$--maps)
$$P_{{\beta}{\gamma}}^{{..}{\alpha}}(x)={\psi}_{({\beta}}{\delta}^
{\alpha}_{{\gamma})}, \eqno (16)$$
where ${\delta}^{\alpha}_{\beta}$ is Kronecker symbol and ${\psi}_{\beta}=
{\psi}_{\beta}(x)$ is the covariant vector fields ;

--for ${\pi}_{(1)}$--maps $P^{{..}{\alpha}}_{{\beta}{\gamma}}(x)$ is the
solution of equations
$$ 3D_{\alpha}P^{{..}{\tau}}_{{\beta}{\gamma}}=
2R^{{.}{\tau}}_{({\beta}{.}{\gamma}){\alpha}}-
2{\underline R}^{{.}{\tau}}_{({\beta}{.}{\gamma}{\alpha})}+
6b_{({\alpha}}P^{{..}{\tau}}_{{\beta}{\gamma})}+
6a_{({\alpha}{\beta}}{\delta}^{\tau}_{{\gamma})};  \eqno (17)$$

--for ${\pi}_{(2)}$--maps
$$P_{{\alpha}{\beta}}^{{..}{\tau}}=
2{\phi}_{({\alpha}}{\delta}^{\tau}_{{\beta})}+
2{\sigma}_{({\alpha}}F^{\tau}_{{\beta})}, \eqno (18) $$
where $F^{\alpha}_{\beta}=F^{\alpha}_{\beta}(x)$
satisfies conditions $F_{\alpha}^{\beta}=F^{\alpha}_{\delta}=e{\delta}^{\beta}_{\delta},
  (e={\pm}1),$
$${\partial}_{[{\beta}}F^{\tau}_{{\gamma}]}F^{\gamma}_{\lambda}-
{\partial}_{[{\lambda}}F^{\tau}_{{\gamma}]}F^{\gamma}_{\beta}=0 \eqno (19) $$
($[\quad]$ denotes antisymmetrization) and solves equations
$$ D_{\beta}F^{\tau}_{\alpha}+
{\mu}_{\gamma}F^{\gamma}_{\alpha}{\delta}^{\tau}_{\beta}-
{\mu}_{\alpha}F^{\tau}_{\beta}=0 \eqno (20) $$
for a covariant vector field ${\mu}_{\gamma}={\mu}_{\gamma}(x);$

--for ${\pi}_{(3)}$--maps
$$P^{{..}{\tau}}_{{\gamma}{\delta}}=
2{\psi}_{({\gamma}}{\delta}^{\tau}_{{\delta})}+
{\sigma}_{{\gamma}{\delta}}{\varphi}^{\tau} , \eqno (21) $$
where the contravariant vector field ${\varphi}^{\tau}={\varphi}^{\tau}(x)$
satisfies equations
$$D_{\alpha}{\varphi}^{\beta}={\nu}{\delta}^{\beta}_{\alpha}+{\mu}_{\alpha}
{\varphi}^{\beta}, \eqno (22) $$
for some scalar,  ${\nu}={\nu}(x), $ covariant vector, ${\mu}_{\gamma}=
{\mu}_{\gamma}(x)$ and ${\varphi}_{\nu}={\varphi}_{\nu}(x)$ symmetric tensor
${\sigma}_{{\alpha}{\beta}}={\sigma}_{{\alpha}{\beta}}(x)$ fields.}
\vskip 5pt

We emphasize  that for ${\varphi}_{\mu}={{{\partial}{\varphi}} \over
{\partial}x^{\mu}}=g_{{\mu}{\nu}}{\varphi}^{\nu}$ and
$ {\sigma}_{{\alpha}{\beta}}(x)=g_{{\alpha}{\beta}}(x)$ we obtain
a particular case of conformal  maps,
${\pi}_{{3}(C)} : {\underline g}_{{\mu}{\nu}}=e^{2{\varphi}}g_{{\mu}{\nu}}$
(the so called concircular maps [15] ).
\vskip 5pt

{\it 2.3. Invariant criterions for ng--flat spaces.}
\vskip 5pt

{\sl Definition 4.} {\it (Pseudo)--Riemannian space $V$ is ng--flat
if it admits a map $ng : V {\to} {\underline M}$ } .

We shall consider four classes of na-flat spaces denoted
respectively as ${\pi}_{(i)}$--flat spaces, where $(i)=((0),(1),(2),(3))$.

It is significant that na--maps are characterized by corresponding invariant
conditions for values being similar to Thomas parameters and Weyl tensor
(the invariants for conformal maps [16]). Here  we present the
criterions for a space $V$ to be ng--flat .
\vskip 5pt

{\sl Proposition 1 :}
{\it For ng--flat spaces there are satisfied the next conditions :

--for ${\pi}_{(0)}$--spaces

$$ W^{{.}{\beta}}_{{\alpha}{.}{\gamma}{\delta}}=
R^{{.}{\beta}}_{{\alpha}{.}{\gamma}{\delta}}-{2 \over 3}R_{{\alpha}[{\gamma}}
{\delta}^{\beta}_{{\delta}]}=0; \eqno (23) $$

--for ${\pi}_{(1)}$--spaces
$$ 3D_{\gamma}P^{{..}{\tau}}_{{\alpha}{\beta}}=
2R^{{.}{\tau}}_{({\alpha}{\beta}){\gamma}}+
6b_{({\alpha}}P^{\tau}_{{.}{\beta}{\gamma})}+
6a_{({\alpha}{\beta}}{\delta}^{\tau}_{{\gamma})}; \eqno (24) $$

--for  ${\pi}_{(2)}$-spaces
$${\hat W}^{{.}{\tau}}_{{\alpha}{.}{\beta}{\delta}}=
{\hat R}^{{.}{\tau}}_{{\alpha}{.}{\beta}{\gamma}}+
{2 \over 5}{\hat R}_{[{\beta}{\gamma}]}{\delta}^{\tau}_{\alpha}-
{2 \over 15}[4{\hat R}_{{\alpha}[{\beta}}{\delta}^{\tau}_{{\gamma}]}+
{\delta}^{\tau}_{[{\gamma}}{\hat R}_{{\beta}]{\alpha}}]=0, \eqno (25)$$
where
$$ {\hat R}_{{\alpha}{.}{\beta}{\gamma}}^{{.}{\tau}}=-
R_{{\alpha}{.}{\beta}{\gamma}}^{{.}{\tau}}+
eF^{\tau}_{\delta}(F^{\lambda}_{\delta}R^{{.}{\delta}}_{{\lambda}{.}{\beta}{\gamma}}+
2R^{{.}{\delta}}_{{\lambda}{.}{\alpha}[{\gamma}}F^{\lambda}_{{\beta}]}-
2D_{[{\gamma}}F^{\delta}_{{\beta}]}-$$
$$2eF^{\lambda}_{\epsilon}D_{({\alpha}}F^{\epsilon}_{{\beta})}D_{\lambda}F^{\delta}
_{\gamma}+
2eF^{\lambda}_{\epsilon}D_{({\alpha}}F^{\epsilon}_{{\gamma})}D_{\lambda}
F^{\delta}_{{\beta})}) \eqno (26) $$
and
$${\hat R}_{{\epsilon}{\tau}}=-
R_{{\epsilon}{\tau}}+eF^{\delta}_{\alpha}(F^{\beta}_{\epsilon}R^{{.}{\alpha}}_
{{\beta}{.}{\tau}{\delta}}+
F^{\beta}_{\tau}R^{{.}{\alpha}}_{{\beta}{.}{\epsilon}{\delta}}-
2D_{\epsilon}D_{[{\delta}}F^{\alpha}_{{\tau}]}-
2eF^{\gamma}_{\beta}D_{({\epsilon}}F^{\beta}_{{\tau})}
D_{\gamma}F^{\alpha}_{\delta}+$$
$$eF^{\gamma}_{\beta}D_{({\epsilon}}F^{\beta}_{{\delta})}D_{\gamma}F^{\alpha}_
{\tau}); \eqno (27) $$}

{\it --for ${\pi}_{(3)}$--spaces
$$R_{{\beta}{\alpha}{\gamma}{\delta}}=A(g_{{\alpha}{\gamma}}g_{{\beta}{\delta}}
-
g_{{\alpha}{\delta}}g_{{\beta}{\gamma}})+
eB[{\varphi}_{\beta}(g_{{\alpha}{\gamma}}{\varphi}_{\delta}-
g_{{\alpha}{\delta}}{\varphi}_{\gamma})-
{\varphi}_{\alpha}(g_{{\beta}{\gamma}}{\varphi}_{\delta}-
g_{{\beta}{\delta}}{\varphi}_{\gamma})], \eqno (28) $$
where $A=-{{1}\over{2}}[{{R}\over{3}}-2({{d{\nu}}\over{d{\varphi}}}
e{\nu}^{2})],$ and
$ A+B+({d\nu \over d\varphi}+e{\nu}^{2})=0,  e={\pm}1, $
for some gradient vector ${\varphi}_{\alpha}={d\varphi \over dx^{\alpha}}$
and  scalar ${\nu}({\varphi}) $} fields.
\vskip 5pt

We note that from (28) one follows this expression for the Ricci tensor of
${\pi}_{3}$--
flat spaces:
$$R_{{\alpha}{\beta}}=[{{R} \over {3}}-({{d{\nu}} \over {d{\varphi}}}+
e{\nu}^{2})]g_{{\alpha}{\beta}}-
[{{R} \over {3}}-4({{d{\nu}} \over
{d{\varphi}}}+e{\varphi}^{2})]e{\varphi}_{\alpha}
{\varphi}_{\beta}.  \eqno (30)$$
\vskip 5pt

{\it 2.4. The integrability conditions for ng--maps equations.}
All presented in this
paper
basic equations for ng--maps ( equations (17), (18) and (20)--(22)) are
systems of first order
partial differential equations with algebraic constraints of type (19).
The integrability conditions for ng--map equations have been studied in [4]
and, in the language of Pffaf systems [17-- 20], in [5, 6, 13].
The most important conclusion made in the just mentioned works is that we can
always
verify, by using algebraic methods, if a given system of ng--map equations on
$V$ is, or not, integrable for maps to Minkowski space. Let illustrate this
for maps
${\pi}_{1} : V {\to} {\underline M}$~ specified by equations
$$ 3(D_{\gamma}P^{{..}{\tau}}_{{\alpha}{\beta}}+
P^{{..}{\epsilon}}_{{\alpha}{\beta}}P^{{..}{\tau}}_{{\gamma}{\epsilon}})=
2R^{{.}{\tau}}_{({\alpha}{.}{\beta}){\gamma}}+
6a_{({\alpha}{\beta}}{\delta}^{\tau}_{{\gamma})} \eqno (31) $$
(these equations can be obtained from (14) by using auxiliary
${\pi}_{0}$--maps ,
$ {{\overline V} \buildrel \pi \over \to}  V {\to} {\underline M}$, see [4]).
The integrability conditions of (31) can be written as
$$ S^{\tau}_{{.}{\alpha}{\beta}{\gamma}{\epsilon}}=
6(D_{\epsilon}a_{({\alpha}{\beta}}
{\delta}^{\tau}_{{\gamma})}-
D_{\gamma}a_{({\alpha}{\beta}}{\delta}^{\tau}_{{\epsilon})}), \eqno (32)$$
where
$$S^{\tau}_{{.}{\alpha}{\beta}{\gamma}{\epsilon}}=-
D_{\epsilon}R^{{.}{\tau}}_{({\alpha}{.}{\beta}){\gamma}}+
D_{\gamma}R^{{.}{\tau}}_{({\alpha}{\beta}){\epsilon}}+
6P_{{\lambda}({\alpha}}^{{.}{\tau}}R_{{\beta}).{\gamma}{\epsilon}}^{{.}{\lambda}}+
4R_{({\alpha}{.}{\beta})[{\epsilon}}^{{.}{\lambda}}
P_{{\gamma}]{\lambda}}^{{.}{\tau}}+$$
$$6P^{{.}{\tau}}_{{\gamma}{\lambda}}{\delta}^
{\lambda}_{({\alpha}}a_{{\beta}{\epsilon})}-
6P^{{.}{\tau}}_{{\epsilon}{\lambda}}{\delta}^{\lambda}_{({\alpha}}
a_{{\beta}{\gamma})} \eqno (33)$$
From (32) we obtain
$$ 3D_{\gamma}a_{{\alpha}{\beta}}={{1} \over 3}S^{\epsilon}
_{{.}{\epsilon}({\alpha}{\beta}){\gamma}}-
S^{\epsilon}_{{.}{\alpha}{\beta}{\gamma}{\epsilon}} \eqno(34)$$

Equations (31) and (34) forms a Cauchy system of first order
partial differential equations (with coefficients given by geometrical
values such as connection and curvature  on $V )$
for unknown variables $P^{{..}{\alpha}}_{{\beta}{\gamma}}$ and
$a_{{\alpha}{\beta}}$.

The first set of integrability conditions for the system of equations
(31) and (34) can be found from (32) by excluding covariant derivations
of $a_{{\alpha}{\beta}}$ according to (34). In result we obtain integrability
conditions being linear equations for deformation tensor $P^{{..}{\gamma}}
_{{\alpha}{\beta}}$. Introducing the second, third and so on sets of integrability
conditions, we are, in general, able to clarify the question of existence of
solutions of system (31) and (34).

We note that in a similar manner we can
construct sets of integrability conditions for ${\pi}_{(2)}$--, (20), and
${\pi}_{(3)}$--map, (22), equations.
\vskip 5pt
{\it 2.5. Spinor formalism and the ng--map theory.}
This question is studied in details in [13, 14] by introducing deformations
by ng--maps of spin coefficients (7) in spinor covariant derivation operator
(8). Using ${\sigma}$--coefficients we can transform basic ng--map equations
(18), (20) and (22) and flat projectivity conditions (23), (24) and (28)
into spinor form. We omit these considerations here. For our purposes
it is important the fact that for every deformation of spin coefficients
$^{\star}{\gamma}_{{\mu}B}^{{..}{C}}(x)$   (see splitting (7))
we can define corresponding deformation tensor (see expressions (5)--(7)),
$$P_{BB^{\prime}CC^{\prime}}^{{....}AA^{\prime}}=
 ^{\star}{\gamma}_{BB^{\prime}C}^{{...}{A}}{\epsilon}_{C^{\prime}}^{{.}A^{\prime}}+
^{\star}{\overline {\gamma}}_{BB^{\prime}}^{{...}A^{\prime}}
{\epsilon}_{C}^{{.}{A}}.  \eqno (35) $$
Putting (35), for example, into (33) we obtain a system of algebraic equations,
if necessary in spinor variables with a spinor representation of curvature
and deformation parameters, which permits us to answer the question if,
the given deformation of spin coefficients generates,  or not, a map
${\pi}_{(1)}: V \to {\underline M}.$

Finally, in this section, we note
that every curved space $V$, if corresponding conditions on differentiability
of components of metric, connection and curvature on $V$ are satisfied, admits
a finite chain of ng--maps, i.e. a nc--transform, to Minkowski space
${\underline M}$ [5--9, 13]. So, it is possible a new classification
of curved spaces in terms of minimal chains of ng--maps characterized by
corresponding sets of invariant conditions of type (23)--(25) and (28). This
ng--map classification of curved  spaces
differs from the well known Petrov's algebraic classification [21].

\section{ Nearly conformal twistors}

The purpose of this section is to define twistors on ng--flat spaces.
\vskip 6pt

{\it 3.1 Spinor equations for massless fields with spin ${m \over 2} (m=0, 1, 2, {\ldots})$
and twistor equations .}
Let spinor ${\phi}_{{AB}{\cdots}L}$ has $m$ indices and is symmetric:
$${\phi}_{{AB}{\cdots}L}={\phi}_{(AB{\cdots}L)}. \eqno (36)$$
The dynamic equations for massless spin ${m \over 2}$ field are written as
$$D^{AA^{\prime}}{\phi}_{{AB}{\cdots}L}=0. \eqno (37)$$
The compatibility conditions [22, 23, 1] of equations (37) for
uncharged spinor field (36) can be written as
$$(m-2){\phi}_{{ABM}(C{\ldots}K}{\Psi}_{L)}^{.ABM}=0,  \eqno (38)$$
   where ${\Psi}_{LABM}$ is the Weyl spinor on space $V$.

Because on generic curved spaces conditions (38) are not satisfied, there is an essential
difficulty in definition of physical fields (36) as solutions of equations (37).

The same difficulty  appears and for twistor equations (10), rewritten
on space $V$ :
$$D^{(A}_{A^{\prime}}{\omega}^{B)}=0, \eqno (39)$$
with compatibility conditions
$$D^{A^{\prime}(C}D_{A^{\prime}}^{A}{\omega}^{B)}=-
{\Psi}^{CA.B}_{..D}{\omega}^{D}.$$
\vskip 6pt
{\it 3.2. Systems of first order partial differential equations.} Mentioned
field, (37), and twistor, (10), equations are systems of first order
partial differential equations. We shall study the general properties of such
systems of equations by using methods of  the geometrical theory of
differential
equations  [17--20].

Let consider, in general form, a system of first order partial differential
equations on space $V^{n}, dim  V^{n}=n$~,
$$ f_{s}=(x^{1}, {\ldots}, x^{n}, y^{1}, {\ldots}, y^{r},
{{{\partial}y^{1}} \over {{\partial}x^{1}}}, {\ldots},
{{{\partial}y^{r}} \over {{\partial}x^{n}}})=0, \eqno (40)$$
where $x^{1}, {\ldots}, x^{n}$ are independent variables,
    $y^{1}, {\ldots}, y^{r}$ are unknown functions and
$s=1, 2, {\ldots}, q.$
Introducing new unknown variables (functions)
$$p_{\dot {\alpha}}^{\dot a}=
{{{\partial}y^{\dot a}} \over {\partial}x^{\dot {\alpha}}}, \quad ({\dot a}=
1, {\ldots}, r; { \dot \alpha}=1, {\ldots}, n), $$
we reduce equations (40) to a Pffaf system
$${\theta}^{\dot a}=dy^{\dot a}-p_{\dot {\alpha}}^{\dot a}dx^{\dot
{\alpha}}=0,
\eqno (41)$$
where variables $p^{\dot a}_{\dot {\alpha}}$ satisfy finite relations
$$ f_{s}(x^{\dot {\alpha}}, y^{\dot a}, p^{\dot a}_{\dot {\alpha}})=0. \eqno
(42)$$
Solving (41) on $q$ independent values $p^{\hat a}=\lbrace
p^{\dot a}_{\dot {\alpha}} \rbrace$
and putting them into (41) we obtain a system of $r$ Pffaf equations on
${\hat r}=r+nr-q$ unknown functions on independent variables
$x^{\dot {\alpha}}$
(differentials $dx^{\dot {\alpha}}$ have the role of distinguished variables).

Let ${\dot U}$ be a open region locally isomorphic to ${\bf R}^{{\hat r}+n}.$
We write the new  Pffaf system as
$${\theta}^{\dot A}={\theta}^{\dot A}_{\hat a}(x^{\hat {\alpha}}, p^{\hat a})
dp^{\hat a}+b^{\dot A}_{\dot a}(x^{\dot {\alpha}},
p^{\hat b})dx^{\dot a}=0 \eqno (43)$$
${\dot A}=1, 2, {\ldots}, r; {\hat a}=1, 2, {\ldots}, r+rn-q).$
Equations ( 42) are linearly independent if
$rang {\Vert}{\theta}^{\dot A}_
{\hat a} {\Vert}{\leq} r$ in every point $x^{\dot {\alpha}}$ of open region
$U{\subset} V^{n}$. We mention that integral varieties $I_{\theta}$ of the
system (43) should be defined from equations
(the closure of (43)):
$${\theta}^{\dot A}=0; $$
$$D{\theta}^{\dot A}=0 , \eqno (44) $$
where quadratic exteriors forms are written as
$$D{\theta}^{\dot A}=a^{\dot A}_{{\hat a}{\hat b}}dp^{\hat a}{\land}
dp^{\hat b}+
C^{\dot A}_{{.}{\hat a}{\dot {\alpha}}}dp^{\hat a}{\land}dx^{a}+
b^{\dot A}_{{\dot a}{\dot b}}dx^{\dot a}{\land}dx^{\hat b}$$

If a solution
$$y^{\dot a}=y^{\dot a}(x^{{\dot \alpha}})  \eqno (45) $$
of equations (40) (or equivalently (41) and (42)) is found, it must satisfy
integrability conditions
$${{{\partial}y^{\dot a}} \over {\partial}y^{\dot {\alpha}}
{\partial}x^{\dot {\beta}}}=
  {{\partial}y^{\dot a} \over {\partial}x^{\dot {\beta}}
{\partial}y^{\dot {\alpha}}},$$
or, equivalently,
$${\partial}_{\dot {\beta}}p_{\dot {\alpha}}^{\dot a}=
{\partial}_{\dot {\alpha}}
p^{\dot a}_{\dot {\beta}}, \eqno (46)$$
i.e. if equations (40) are compatible, the Pffaf system (41) can be reduced
to total differential relations,
$$ d {\lambda}^{\dot a}=dy^{\dot a}-p^{\dot a}_{\dot {\alpha}}dx^{\dot
{\alpha}}
=0.$$
In this case solution (45) should be obtained from relations
$${\lambda}^{\dot a}(y^{\dot b}, x^{\dot {\alpha}}) = C^{\dot a},\  C^{\dot a}=
const, $$
where $rang{\Vert}{{\partial}{\lambda}^{\dot a} \over {\partial}y^{\dot b}}
{\Vert}{\neq}0. $
If conditions (46) are not satisfied, one tries to solve equation (40)
by introducing unknown functions
$${\tilde p}^{\dot a}_{\dot {\alpha}}=p^{\dot a}_{\dot {\alpha}}-
{{\partial}{\zeta}^{\dot a} \over {\partial}x^{\dot {\alpha}}} \eqno (47)$$
and considering a new Pffaf system
$${\tilde {\theta}}^{\dot a}=dy^{\dot a}-
{\tilde p}^{\dot a}_{\dot {\alpha}}dx^{\dot {\alpha}}=0, \eqno (48) $$
where $ {\tilde f}_{s}(x^{\dot {\alpha}}, y^{\dot a}), {\tilde p}^{\dot
a}_{\dot {\alpha}}= 0.$
To obtain a total differential relation we multiply (48) on a
nondegenerated
matrix function
${\mu}^{{.}{\dot b}}_{\dot a}(x^{\dot {\alpha}}, y^{\dot a}):$
$$ d{\tilde {\lambda}}^{\dot b}={\mu}_{\dot a}^{{.}\dot b}dy^{\dot a}-
{\mu}^{{.}{\dot b}}_{\dot a}{\tilde p}^{\dot a}_{\dot {\alpha}}dx^{\dot
{\alpha}}=0. \eqno (49)$$
Integrating system (49) we obtain relations
$${\tilde {\lambda}}^{\dot b}(x^{\dot {\alpha}}, y^{\dot a})={\tilde C}^{\dot
b},
{\tilde C}^{\dot b}=const, $$
from which the solution $y^{\dot a}=y^{\dot a}(x^{\dot {\alpha}})$ of
equations (40) can be found an explicit form.
We note that if deformation functions ${\zeta}^{\dot a}$ from (46) and
integrating matrix ${\mu}_{\dot a}^{{.}{\dot b}}$ from
(48) exist, the Pffaf system (41) can be expressed as
$${\theta}^{\dot a}=dy^{\dot a}-p^{\dot a}_{\dot {\alpha}}dx^{\dot {\alpha}}=
d{\zeta}^{\dot a}+{({\mu}^{-1})}^{{.}{\dot a}}_{\dot b}d{\tilde {\lambda}}
^{\dot b}, $$
where ${({\mu}^{-1})}^{{.}{a}}_{\dot b}$ is inverse to ${\mu}^{{.}{\dot b}}_
{\dot a}$. As particular cases we can consider the trivial integrating matrix,
${\mu}_{\dot a}^{{.}{\dot b}}={\delta}^{\dot b}_{\dot a},$ and (or)
vanishing of deformation when ${\zeta}^{\dot a}=const.$

We also emphasize that introducing new, deformed, variables (47) and matrix
${\mu}^{{.}{\dot b}}_{\dot a}$ into relations (42) we obtain new finite
relations ${\tilde f}_{s}(x^{\dot {\alpha}}, y^{\dot a}, p^{\tilde a}_{\tilde
{\alpha}})=0$
which, as a matter of principle, reflects the deformation of first order
partial differential equations (40) into another one ( really for every
given system (40) one exists an infinite number of deformations (46)
and integrating matrices ${\mu}^{{.}{\dot b}}_{\dot a}$.
Considering equations (40) as a fundamental physical equations
of type (37), or (39), one arises the problem of definition a unique
deformation,
motivated from physical and geometrical point of view, transforming mentioned,
in general uncompatible on curved spaces, equations into compatible ones,
on some auxiliary spaces.
\vskip 6 pt

{\it 3.3. Ng--deformations of twistor equations .} As was shown   twistor
equations happen to be
uncompatible on a given curved space $V$. Our aim is to formulate an algorithm
of transporting mentioned differential equations, by using ng--maps,
 from the space $V$ to another one, ${\underline V}$, on which
compatibility conditions (11) will be satisfied. In this case we
extend our system of twistor equations by introducing into consideration
additional variables (deformation parameters (15), $a_{{\alpha}{\beta}}$
and $b_{\alpha}$, and deformation of connection,
$P^{{..}{\alpha}}_{{\beta}{\gamma}},$
and supplement the initial twistor system on $V$ with a system of basic ng--map
equation to ${\underline V}.$ Let illustrate this construction for
twistor equations (39) rewritten, as a Pffaf system  (41),
$$d{\omega}^{A}-p^{ C^{\prime}CA}dx_{CC^{\prime}}=0, \eqno (50) $$
where unknown functions
$$p^{C^{\prime}CA}={{\partial}{\omega}^{A} \over {\partial}x_{CC^{\prime}}}$$

 satisfy finite relations, of type (42),
$$ p^{A^{\prime}(AB)}+{\gamma}^{A^{\prime}(A{.}B)}_{{..D}}{\omega}^{D}=0,
\eqno (51) $$
${\gamma}^{A^{\prime}{A}.B}_{{..}D}$ are spin--coefficients on space $V$.
Twistor equations (39), and their associated Pffaf system (50) and (51),
are incompatible on generic curved space--time $V$. We suggest to extend
 the mentioned system of equations by considering a new  system
  of differential equations
$$D^{A^{\prime}(A}{\tilde {\omega}}^{B)}={\Lambda}^{A^{\prime}(A.B)}_{{..}D}
{\tilde {\omega}}^{D}, \eqno (52) $$
or equivalently, a new, associated to (52), Pffaf system
$$d{\tilde {\omega}}^{A}-{\tilde p}^{C^{\prime}CA}dx_{CC^{\prime}}=0, \eqno
(53)$$
where unknown functions ${\tilde p}^{C^{\prime}CA}=
{{\partial}{\tilde {\omega}}^{A} \over {\partial}x_{CC^{\prime}}}$
must satisfy relations
$${\tilde p}^{A^{\prime}(AB)}+{\gamma}^{A^{\prime}(A.B)}_{..D}
{\tilde {\omega}}^{D}+{\Lambda}^{A^{\prime}(A.B)}_{..D}
{\tilde {\omega}}^{D}=0. \eqno (54)$$

Spinors ${\Lambda}^{A^{\prime}A.B}_{..D}$ from (52)
can be considered as obtained in result of a deformation of type (47) and
a multiplication on integrating matrix as in (49).
Introducing  ng--maps we identify
${\Lambda}$--spinors with deformation of spin coefficients
${^{\star}{\gamma}}_{{\mu}B}^{{..}{C}}$ (see relations (7) ):
$$ {\Lambda}_{A^{\prime}AD}^{{...}B}={^{.}{\gamma}}_{AA^{\prime}D}^{{...}B}.
\eqno (55)$$

{\sl Proposition 2.} {\it Deformed twistor equations (52) ( and  associated
twistor Pffaf system (53) and (54)) are compatible if spinors (55) solve one
of the ng--map equations ((16), (17), (18)--(20) and (21)--(22)) and
satisfy one of the corresponding ng--flat criterions ((23), (24), (25) and
(28)).}
\vskip 5pt

{\bf Proof}. Defining new spin coefficients
$${\tilde {\gamma}}_{AA^{\prime}D}^{...B}=
         {\gamma}_{AA^{\prime}D}^{...B}+
        {\Lambda}_{AA^{\prime}D}^{...B},$$
which according to our proposition become trivial (with vanishing curvature)
spin--coefficients on flat space ${\tilde M}.$ In this case equations (52)
can be written as
$${\tilde D}^{A^{\prime}(A}{\tilde {\omega}}^{B)}=0 . \eqno (56)$$
Equations (56) are compatible because on the flat space ${\tilde M}$  the
Weyl tensor vanishes (see relations (11), (12)). The proposition is proved.
${\diamond}$
\vskip 5pt

Instead of ng--maps we can consider chains   of ng--maps (nc--transforms)
$nc : V {\to} {\underline M}.$ Nc--twistors are defined as solutions of
deformed twistor equations with ${\Lambda}$--spinor, being a superposition
of spin tensors,
$${\Lambda}_{A^{\prime}AB}^{...D}=
{^{\star}{_{(1)}{\gamma}}}_{AA^{\prime}B}^{...D}+
{^{\star}{_{(2)}{\gamma}}}_{AA^{\prime}B}^{...D}+
{\cdots}+
{^{\star}{_{(k)}{\gamma}}}_{AA^{\prime}B}^{...D}, \eqno (57)$$
associated to a finite chain of ng--maps.
In a particular case when (57) reduces to (55) we obtain ng--twistors.
\vskip 6pt

{\it 3.4. Ng--images of twistors.} On flat space ${\underline M}$ twistors
are defined as  a pair of spinors, ${\underline Z}^{\alpha}=
({\underline {\omega}}^{A}, {\underline {\pi}}_{A}),$
where
$$ {\underline {\omega}}^{A}={^{(0)}{\omega}}^{A}-ix^{AB^{\prime}}
{^{(0)}{\pi}}_{B^{\prime}}, \quad  {\underline {\pi}}_{A^{\prime}}=
{^{(0)}{\pi}}_{A^{\prime}}=const, $$
is the general solution of twistor equations
$${\underline D}^{A^{\prime}(A}{\underline {\omega}}^{B)}=0.$$

Nc--twistors on space $V$, being nc--coimage of space ${\underline M}$,
for a given map $nc: V {\to} {\underline M}, $ are defined as a pair
of spinors $Z^{\alpha}=({\omega}^{A}, {\pi}_{A^{\prime}}), $ where
${\omega}^{A}$ is the general solution of nc--twistor equations
$$ D^{A^{\prime}(A}{\omega}^{B)}={\Lambda}^{A^{\prime}(A.B)}
_{..D}{\omega}^{D}, \eqno (58)$$
with ${\Lambda}$--spinors defined from (57).
For a local common spinor coordinate system on spaces under consideration
we can write ${\omega}^A={\underline {\omega}}^A$ and define the second spinor
${\pi}_{A^{\prime}}$ as ${\pi}_{A^{\prime}}={i \over
2}D_{AA^{\prime}}{\omega}^{A}.$
Taking into account that ${\underline {\pi}}_{A^{\prime}}=
{i \over 2}{\underline D}_{AA^{\prime}}{\omega}^A, $
we have
$${\pi}_{A^{\prime}}
=^{(0)}{\pi}_{A^{\prime}}-{i \over 2}{\Lambda}_{AA^{\prime}C}^{...A}
{\omega}^{C}. \eqno (59)$$

In a similar manner we can define dual nc--twistors on $V$ as pairs of spinors
$W_{\alpha}=({\lambda}_A, {\mu}^{A^{\prime}})$, where
$$ {\lambda}_A={^{(0)}{\lambda}}_{A}+
{i \over
2}{\Lambda}_{AC^{\prime}A^{\prime}}^{...C^{\prime}}{\mu}^{A^{\prime}}, \quad
{^{(0)}{\lambda}}_{A}=const,$$
and ${\mu}^{A^{\prime}}={^{(0)}{\mu}}^{A^{\prime}}+
ix^{AA^{\prime}}{^{(0)}{\lambda}}_{A}$
is the general solution of dual nc--twistor equation
$$D^{A(A^{\prime}}{\mu}^{B^{\prime})}=
{\Lambda}^{A(A^{\prime}{.}B^{\prime})}_{{.}{.}D^{\prime}}{\mu}^{D^{\prime}},
$$
where spinors
${\Lambda}_{AA^{\prime}D^{\prime}}^{{.}{.}{.}C^{\prime}}$ are defined as a
superposition
of ng--transformation as (57).

We end this section by concerning the question of geometrical
interpretation of nc--twistors . To an isotropic twistor
${\underline Z}^{\alpha}=({\underline {\omega}}^{A},
{\pi}_{A^{\prime}})\not=0,
{\underline Z}^{\alpha}{\underline {\overline Z}}_{\alpha}=0 \quad
({\underline {\overline Z}}_{\alpha}=({\underline {\overline {\pi}}}_{A},
{\underline {\overline {\omega}}}^{A^{\prime}})$ denotes complex conjugation
of ${\underline Z}^{\alpha})$ one associates [2] an isotropic line
on space ${\underline M}$:
$$x^a = ^{(0)}x^a + {\eta}{{\underline \xi}}^a,
 \quad {\eta}_1 <{\eta} < {\eta}_2,
\eqno (60)$$
where $ ^{(0)}x^a =
{(i\  ^{(0)}{{\underline {\overline \omega }}}^{B^{\prime}}\
 ^{(0)}{{\underline \pi}}_{B^{\prime}})}^{-1}\  ^{(0)}{{\underline \omega}}^
{A}\   ^{(0)}{{\underline {\overline {\omega}}}}^
{A^{\prime}},\
{\underline {\xi}}^a
=\  ^{(0)}{\underline {\overline \pi}}^{A} \
 ^{(0)}{\underline \pi }^{A^{\prime}}.$
The
nc--coimages of ${\underline Z}^{\alpha}$ and
${\underline {\overline Z}}^{\alpha}$ on the space $V$, defined as
$Z^{\alpha} = ({\omega}^A, {\pi}_{A^{\prime}})$ and, respectively, as
${\overline Z}_{\alpha}=({\overline {\pi}}_{A}, {\overline
{\omega}}^{A^{\prime}}),$
where
$$ {\omega}^{A}={\underline {\omega}}^A= ^{(0)}{\underline {\omega}}^A-
ix^{A{A^{\prime}}}{ ^{(0)}{\underline {\pi}}}_{A^{\prime}}, \quad
{\pi}_{A^{\prime}}= ^{(0)}{\pi}_{A^{\prime}}-
{i \over 2}{\Lambda}_{A{A^{\prime}}C}^{...A}{\omega}^{C}, \eqno (61) $$
and ${\overline {\pi}}_{A}= ^{(0)}{\overline {\pi}}_A-
{i \over 2}{\overline {\Lambda}}_{{A^{\prime}}A{C^{\prime}}}^{...{A^{\prime}}}
{\overline {\omega}}^{C^{\prime}},  {\overline {\omega}}^{A^{\prime}} =
{\overline {\underline {\omega}}}^{A^{\prime}}.$
Using spinors (61) we can verify that
  $$ Z^{\alpha}{\overline Z}_{\alpha} \neq 0,  \eqno (62) $$
i.e. a nc--twistor $Z^{\alpha}$,  defined by an isotropic twistor
${\underline Z}^{\alpha}$ is not  isotropic.
So,  nc--twistors on space $V$ parametrize a class of curves
on this space, as theirs nc--images (of type (60)) on flat
space ${\underline M}$, but, in distinction to usual isotropic twistors,
to  an nc--twistor one must associate a  nearly geodesic on $V$
being a corresponding nc-deformation of a isotropic line in flat space.
 Really, the nc--image  of isotropic line (60) on ${\underline M}$
is a curve $l$ on $V$ ( because of equality (62)) with tangent
vector
${\xi}^a={\pi}^A{\overline {\pi}}^{A^{\prime}}$ and complementary
2--dimensional
distribution defined, for example, by
${\xi}^{a}_{(1)}={\xi}^{b}D_{b}{\xi}^{a}$
and ${\xi}_{(2)}^{a}={\xi}^{b}D_{b}{\xi}^{b}_{(1)}, $
where $D_{b}={\underline D}_{b}+{^{\star}{\underline D}}_{b},
^{\star}{\underline D}_b$
is the deformation of connection associated to map $nc: {\underline M} \to V.$

\section{ Deformation of spinor and twistor structures and generation of
curved spaces}
\vskip 5pt

Let on flat space ${\underline M}$ with given primitive spin coefficients
${\underline {\gamma}}_{AA^{\prime}C}^{...B}, $ or connection
${\underline {\Gamma}}_{{\alpha}{\beta}}^{..{\gamma}}, $
( with vanishing curvature) is defined a global twistor structure
as a solution of twistor equations on ${\underline M}.$ Our task is to specify
conditions when a deformed twistor structure, obtained as a solution of
deformed twistor equations (58) with deformation ${\Lambda}$--spinor
of type (55), will generate a ng--flat (pseudo)--Riemannian space--time.

Firstly,  we fix a spinor ${\Lambda}_{A^{\prime}AD}^{...B}$. It is still
not clear if the new connection ${\Gamma}^{..{\alpha}}_{{\beta}{\gamma}},$
defined as to satisfy relations
$${\underline {\Gamma}}_{{\alpha}BB^{\prime}}^{...CC^{\prime}}-
P_{{\alpha}BB^{\prime}}^{...CC^{\prime}}=
{\Gamma}_{{\alpha}BB^{\prime}}^{...CC^{\prime}}, \eqno (63) $$
where
$$P_{{\alpha}BB^{\prime}}^{...CC^{\prime}}=-
{\Lambda}_{{\alpha}B}^{..C}{\epsilon}_{B^{\prime}}^{.C^{\prime}}-
{\overline {\Lambda}}_{{\alpha}B^{\prime}}^{..C^{\prime}}{\epsilon}_{B}^{.C}
\eqno (64) $$
(for simplicity we consider torsionless connections) will generate both
compatible ng--twistor equations (56) and basic ng--map equations
associated to a mutual transform $ng: V \to {\underline M}.$
 We try to give an answer at this question in the following way.
Calculating auxiliary curvature and Ricci tensor for connection (63)
and putting both these expressions and deformation tensor (64) into relations
(32)--(34) we obtain an algebraic system of equations. If this system
is satisfied for some deformation parameters $a_{{\alpha}{\beta}}$ and
$b_{\gamma}$ (see formulas (15)) it is clear that we have obtained a
${\pi}_{(1)}$--flat space $V$.
\vskip 6pt
{\sl Proposition 3} {\it Deformation spinor ${\Lambda}_{AA^{\prime}D}^{...B}$,
and its corresponding deformation tensor $P_{{\beta}{\gamma}}^{..{\alpha}}$
(see (64)), will generate a vacuum Einstein field, if and only if, it is
compatible this system of partial differential equations:
$${\partial}_{{\alpha}}P^{..{\alpha}}_{{\beta}{\gamma}}=
2(b_{({\alpha}}P^{..{\alpha}}_{{\beta}{\gamma})}+a_{({\alpha}{\beta}}{\delta}
^{\alpha}_{{\gamma})})  \eqno (65)$$
($b_{\alpha}$ and $a_{{\alpha}{\beta}}$ are some covariant vector and,
respectively, symmetric
tensor fields).}
\vskip 5pt
{\bf Proof.} We sketch the proof by observing that equations (65) can be
obtained
by contracting indices ${\alpha}$ and $\tau$ in equations (18) written for a
map
${\pi}_{(1)}: {\underline M}, $ where $R_{{\alpha}{\beta}}=0$
and ${\underline R}_{{\alpha}{\beta}{\gamma}{\delta}}=0.$
Of course, to find in explicit form general solutions of equations (65)
is also a difficult task. But we can verify, by solving algebraic equations
(see considerations from subsection 2.4) if equations (65) are, or not,
integrable.$\diamond$
\vskip 5pt

In a similar manner we can analyze the problem of generation
of ${\pi}_{(2)}$--flat and ${\pi}_{(3)}$--flat spaces. Let consider, for
example, ${\pi}_{(2)}$--transforms. In this case we shall parametrize the
deformation spinor (55) in a form as to induce a deformation tensor of type
(18). ${\Lambda}$--spinors should be also chosen as to induce a deformation tensor
(64)
satisfying conditions (25) for ${\pi}_{(2)}$--maps. Calculating
auxiliary curvature and Ricci tensors for connection (63) and putting
both mentioned tensors,  and taking into account basic ${\pi}_{(2)}$--map
equations, into expressions (26) and, respectively, (27) we obtain that
criterion (25) is an algebraic equation on tensors
$R_{{\alpha}{\beta}{\gamma}{\delta}}, R_{{\alpha}{\beta}}, F^{\alpha}_{\beta}$
and  covariant vector field ${\mu}_{\gamma}$.

It is evident that foregoing considerations point to mutual interrelation
between integrable deformations of twistor equations and criterions of
invariance
and integrability of basic equations for ng--maps rather then constitute a
method
of solution of Einstein equation because explicit constructions of metric have
not
been considered in our study. Perhaps, more convenient for the  twistor
treatment
of gravity is the twistor--gauge formulation of gravity on flat nearly
autoparallel
backgrounds [7, 12, 14]. The interrelation between nc--twistors and gauge
gravity is a matter of our further investigations.

Finally we remark that this paper contains a part of results (on definition
of nc-twistors on locally isotropic spaces) communicated at the Coloquium
on Differential Geometry (Debrecen, Hungary, 25-30 July, 1994) [24].
 There we have also
presented some  generalizations on spinor and twistor calculus
for locally anisotropic spaces (which generalize Lagrange and Finsler
spaces) [25,26]. The geometric constructions developed in this paper and in
[12,7] (in the framework of so-called twistor-gauge treatment of
gravity) admit a straightforward extension to locally anisotropic spaces if
we apply  the formalism of locally anisotropic spinors and twistors [27,28]
 and  use gauge like formulations of locally anisotropic gravity
[26,11,10,13,14].

\end{document}